\documentclass[12pt,a4paper,preprint]{article}
\usepackage{multirow} 

\newcommand{\beq}{\begin{equation}} 
\newcommand{\eeq}{\end{equation}} 
\newcommand{\beqa}{\begin{eqnarray}} 
\newcommand{\eeqa}{\end{eqnarray}} 
\newcommand{\mx}{\left[\begin{array}} 
\newcommand{\finmx}{\end{array}\right]} 
\newcommand{\mxp}{\left(\begin{array}} 
\newcommand{\finmxp}{\end{array}\right)} 
\newcommand{\casos}{\left\{\begin{array}} 
\newcommand{\fincasos}{\end{array}\right.} 
\newcommand{\rcasos}{\left.\begin{array}} 
\newcommand{\rfincasos}{\end{array}\right\}}

\def\lsim{\ \rlap{\raise 3pt \hbox{$<$}}{\lower 3pt \hbox{$\sim$}}\ }
\def\gsim{\ \rlap{\raise 3pt \hbox{$>$}}{\lower 3pt \hbox{$\sim$}}\ }
\def\ominus#1{\ \rlap{\raise 0pt \hbox{$#1$}}{\raise 1pt \hbox{$^{^{(-)}}$}}\ }

\begin{document}

\title{
\vspace{-1cm} 
  \begin{flushright}
\normalsize
hep-ph/0412024 \\
UdeA-PE-04/009\\
  \end{flushright}
\bigskip
Supernova Neutrinos and the absolute scale of neutrino masses - a
  Bayesian approach} 

\author{Enrico Nardi\footnote{
\uppercase{W}ork done in collaboration with \uppercase{J. Z}uluaga. 
\uppercase{S}upported in part by \uppercase{COLCIENCIAS}.}\\ 
\small \it 
INFN, Laboratori Nazionali di Frascati, C.P. 13, I00044 Frascati,  Italy \\[-3pt]
\small \it Instituto de F\'\i sica,  Universidad de Antioquia, A.A.{\it 1226}, 
Medell\'\i n, Colombia}

\date{}

\maketitle

\abstract{We apply Bayesian methods to the study of the sensitivity to
neutrino masses of a Galactic supernova neutrino signal. Our procedure
makes use of the full statistics of events and is remarkably
independent of astrophysical assumptions.  Present detectors 
can reach a sensitivity down to $m_\nu \sim 1$ eV.  Future megaton  
detectors can yield up to a factor of two improvement; however, they will not
be competitive with the next generation of tritium $\beta$-decay and
neutrinoless  $2\beta$-decay experiments.}




\section{Introduction}
%
It was realized long time ago that neutrinos from a Supernova
(SN) can provide valuable informations on the neutrino 
masses. 
The basic idea relies on the time-of-flight (tof) delay $\Delta t$
that a neutrino of mass $m_\nu$ and energy $E_\nu$ traveling a
distance $L$ would suffer with respect to a massless particle:
\beq
{\Delta t} = \frac{L}{v} - L 
\approx 
5.1\,\rm ms\;\left(
\frac{L}{10\,\rm kpc}\right)  
\! \left(\frac{10\, \rm
    MeV}{E_\nu}\right)^2  \!\! 
  \! \left(\frac{m_\nu}{1\,\rm eV}\right)^2\,.
\label{eq:delay}
\eeq
Indeed the detection of about two dozens of neutrinos from SN1987 allowed to
set  model independent upper limits at the level of 
$m_{\overline \nu_e}<30\,$eV \cite{Schramm:1987ra}.  
Since SN1987A, several proposal have been put forth to identify the best ways
to measure the neutrino tof delays.  Often, these approaches rely on the
identification of ``timing'' events that are used as benchmarks for measuring
the time lags, as for example the simultaneous emission of gravitational 
waves \cite{gravit} or the abrupt interruption of the neutrino flux due to a further
collapse into a black hole \cite{BH}. The less model dependent limits
achievable with these methods are at the level of $m_{\nu}\lsim 3\,$eV, and
tighter limits are obtained only under specific assumptions.  A different
method to extract informations on the neutrino masses that uses the full
statistics of the signal and does not rely on any benchmark event
was proposed in Ref.~\cite{Nardi:2003pr} and developed in
Ref.~\cite{Nardi:NPB}.  Here we resume the results obtained by applying
this method to determine the sensitivity of the SuperKamiokande (SK) water
\v{C}erenkov and KAMLAND scintillator detectors, and of the planned
experimental facilities Hyper-Kamiokande and LENA.


\section{The method} 
%
In real time detectors, supernova $\bar\nu_e$ are revealed through to the
positrons they produce via charged current interactions, that provides good
energy informations as well.  Each $\bar{\nu_e}$ event corresponds to a pair
of energy and time measurements $ (E_i, t_i)$.  In order to extract the
maximum of information from a high statistics SN neutrino signal, all the
neutrino events have to be used in constructing a suitable statistical
distribution, as for example the Likelihood, that can be
schematically written as: 
\beq
{\mathcal L} \equiv  \prod_{i} {\mathcal L}_i = 
    \prod_{i}\>\big\{
\phi(t_i) \times F(E_i;t_i)\times \sigma(E_i)\big\}\,. 
\label{eq:schematic}
\eeq
${\mathcal L}_i$ represents the contribution to the Likelihood of a single
event with the index $i$ running over the entire set of events, $\sigma(E)$ is
the $\bar\nu_e$ detection cross-section and $F(E;t)$ is the energy spectrum of
the neutrinos, whose time profile can be reconstructed rather accurately
directly from the data \cite{Nardi:2003pr,Nardi:NPB}. The main problem in
constructing the Likelihood (\ref{eq:schematic}) is represented by the
(unknown) time profile of the neutrino flux $\phi(t)$.  We construct a flux
model by requiring that it satisfies some physical requirements and a
criterium of simplicity: {\it i)} the analytical flux function must go to zero
at the origin and at infinity; {\it ii)} it must contain at least two time
scales for the neutrino emission corresponding to the fast rising initial
phase of shock-wave breakout and accretion, and the later Kelvin-Helmholtz
cooling phase; {\it iii)} it must contain the minimum possible number of free
parameters.  The
following model for the flux has all the required behaviors: 
\beq
\phi(t;\lambda) = \frac{ e^{-(t_a/t)^{n_a}}}{[1 + (t/t_c)^{n_p}]^{n_c/n_p}}
\casos{ll}
\sim e^{-(t_a/t)^{n_a}} & (t \to 0) \\
\sim (t_c/t)^{n_c} & (t \to \infty)\,. 
\fincasos
\label{eq:fluxmodel1}
\eeq
The five parameters that on the l.h.s of (\ref{eq:fluxmodel1}) have been
collectively denoted with $\lambda$ are: two time scales $t_a$ for the initial
exponentially fast rising phase and $t_c$ for the cooling phase, two exponents
$n_a$ and $n_c$ that control their specific rates and one additional exponent
$n_p$ that mainly determines the width of the ``plateau'' between the two
phases.  Since we are interested only in the neutrino mass squared $m^2_\nu$,
irrespectively of the particular values of the {\it nuisance parameters}
$\lambda$, starting from the Likelihood (\ref{eq:schematic}) we will need to
evaluate the {\it marginal posterior probability} $p(m^2_\nu|D)$, that is the
probability distribution  for $m^2_\nu$ given the  data
$D$.  This is done by {\it marginalizing} the posterior probability with
respect to the nuisance parameters:
\beq
p(m^2_\nu|D,I)   
 =  N^{-1}\;\int{d \lambda\; {\mathcal L}(D;m^2_\nu,\lambda)\; p(m^2_\nu,\lambda|I)}\,.
\label{eq:marginalization}
\eeq
$p(m^2_\nu,\lambda|I)$, that in
Bayesian language is called the {\it prior probability of the  model},
allows us to take into account any available prior information on the 
parameters $m^2_\nu$ and $\lambda$. We will use
flat priors for all the $\lambda$'s and, to exclude unphysical
values of $m_\nu^2$, a step function
$\Theta(m^2_\nu)=1,\,(0)$ for $m^2_\nu\geq 0,\,(<0)$.

The dependence on $m^2_\nu$ could be directly included in the flux 
(\ref{eq:schematic}) by redefining  the time variable according
to (\ref{eq:delay}).  However, it is more convenient to proceed as
follows: given a test value for the neutrino mass, first the arrival time of
each neutrino is shifted according to its time delay, and then the
Likelihood is computed for the whole time-shifted sample.  Subtilities in the
evaluation of the Likelihood contribution ${\mathcal L}_i(t_i,E_i)$ 
arising from the uncertainty $\Delta E_i$ in the energy
measurement are discussed in Refs.~\cite{Nardi:2003pr,Nardi:NPB}.

We have tested the method applying it to a large set of synthetic Monte Carlo
(MC) neutrino signals, generated according to two different SN models: {\it SN
model I} corresponds to the simulation of the core collapse of a 20~$M_\odot$
star \cite{Woosley:1994ux} carried out by using the Livermore Group
code \cite{Livermore}. In this simulation $\nu_{\mu,\tau}$ opacities were
treated in a simplified way, and this resulted in quite large (and probably
unrealistic) differences in their average energies with respect to
$\bar\nu_e\,$.  {\it SN model II} corresponds to a recent 
hydrodynamic simulation of a 15 $M_\odot$ progenitor star \cite{Raffelt:2003en}
carried out with the Garching group code \cite{GarchingCode}. This simulation
includes a more complete treatment of neutrino opacities and results in a
quite different picture, since the antineutrino spectra do not differ for more
than about 20\%.  In both cases the effects of neutrino oscillations in the SN
mantle have been properly included in the simulations.  Note that
the two types of neutrino spectra of SN model I and II fall close to the two
extremes of the allowed range of possibilities. This gives us confidence that
the results of the method are robust with respect to variations in the
spectral characteristics.


\section{Results} 
%
To test the sensitivity of our method, we have analyzed a large number of
neutrino samples, grouped into different ensembles of about 40 samples
each. For each ensemble we vary in turn the SN model (model I and II), the
SN-earth distance (5, 10, and 15 kpc) and the detection parameters specific
for two operative (SK and KamLAND) and two proposed (Hyper-Kamiokande and
LENA) detectors.  The results of our recent detailed analysis \cite{Nardi:NPB}
are summarized in Table~\ref{tab:results}. In columns 2 and 4 we give the 90\%
c.l.  upper limits that could be put on $m_\nu$ in case its value is too small
to produce any observable delay. In columns 3 and 5 we estimate for which
value of $m_\nu$ the massless neutrino case can be rejected at least in 50\%
of the cases.  These results confirm the claim \cite{Nardi:2003pr} that
detectors presently in operation can reach a sensitivity of about 1~eV, that
is seizable better than present results from tritium $\beta$-decay
experiments, competitive with the most conservative limits from neutrinoless
double $\beta$-decay, less precise but remarkably less dependent from prior
assumptions than cosmological measurements.  However, in spite of a sizeable
improvement, future detectors will not be competitive with the next generation
of tritium $\beta$-decay and neutrinoless double $\beta$ decay experiments.

\begin{table}[ht]
 \setlength{\tabcolsep}{3mm}
\renewcommand{\arraystretch}{1.2}
\begin{center}
\begin{tabular}{lccccc}
  \multicolumn{1}{c}{\vbox{\vskip14pt}\strut } 
& \multicolumn{2}{c}{\bf  MODEL 1} 
&& \multicolumn{2}{c}{\bf  MODEL 2} \\ [3pt] \hline
\multicolumn{1}{l}{\vbox{\vskip15pt}Detector\qquad\qquad\qquad}
& $\overline{m}_{\rm up} \pm \Delta m_{\rm up}$ 
& $\sqrt{m_{\rm min}^2}$ 
&& $\overline{m}_{\rm up} \pm \Delta m_{\rm up}$ 
& $\sqrt{m_{\rm min}^2}$            \\ [3pt] \hline
\vbox{\vskip15pt}\strut 
  a) SK (10 kpc) & $1.0\pm 0.2$ & $1.0$ && $1.1\pm 0.3$ & $1.2$ \\ 
  b) SK (5 kpc) & $-$ & $-$ && $1.1\pm 0.3$ & $1.0$ \\ 
  c) SK (15 kpc) & $-$ & $-$ && $1.6\pm 0.6$ & $1.4$ \\ 
  d) SK+KL (10 kpc)  & $1.0\pm 0.2$ & $0.9$ && $1.1\pm 0.3$ & $1.0$ \\ [5pt] \cline{1-6}
\vbox{\vskip10pt}\strut 
  e) HK (10 kpc) & $0.4\pm 0.1$ & $0.4$ && $0.5\pm 0.1$ & $0.5$ \\ 
  f) LENA (10 kpc)  & $0.9\pm 0.2$ & $0.9$&& $0.9\pm 0.3$ & $0.9$ \\ [1pt] \cline{1-6}
\end{tabular}
\end{center}
\vspace*{-5pt}
\caption{Results for the fits to $m^2_\nu$:  a)-c) SK 
for different SN distances, d) SK plus KamLAND, 
e) Hyper-Kamiokande, f) LENA.  All masses are in eV.}
\label{tab:results}
\vspace*{-3pt}
\end{table}

\vspace*{.1cm}



\end{document}